\documentclass[prb,superscriptaddress,showpacs,letterpaper]{revtex4}
\usepackage{amsmath}
\usepackage{graphicx}
\usepackage[backref,colorlinks,linkcolor=blue,citecolor=blue]{hyperref}


\begin{document}

\title{Derivation of the three-layer model for surface second-harmonic
       generation}
\author{Sean M. Anderson}
    \affiliation{Centro de Investigaciones en \'Optica, 
                Le\'on, Guanajuato, M\'exico}
\author{Bernardo S. Mendoza}\email{bms@cio.mx}
    \affiliation{Centro de Investigaciones en \'Optica, 
                Le\'on, Guanajuato, M\'exico}
\date{\today}

\begin{abstract}
We develop explicit expressions for the surface second-harmonic radiation yield
using the three layer model. We derive expressions that can be applied to
systems without and symmetry considerations, and then reduce them for the (111),
(110), and (100) surface symmetries.
\end{abstract}

\maketitle

%%%%%%%%%%%%%%%%%%%%%%%%%%%%%%%%%%%%%%%%%%%%%%%%%%%%%%%%%%%%%%%%%%%%%%%%%%%%%%%%
%%%%%%%%%%%%%%%%%%%%%%%%%%%%%%%%%%%%%%%%%%%%%%%%%%%%%%%%%%%%%%%%%%%%%%%%%%%%%%%%

\section{Introduction}\label{sec:intro}

Surface second-harmonic generation (SSHG) has been shown to be an effective,
nondestructive and noninvasive probe to study surface and interface
properties.\cite{chenPRL81, shenNAT89, mcgilpOE94, bloembergenAPB99,
mcgilpSRL99, lupkeSSR99, downerPSSA01, downerSIA01} SSHG spectroscopy is now
very cost-effective and popular because it is an efficient method for
characterizing the properties of buried interfaces and nanostructures. The high
surface sensitivity of SSHG spectroscopy is due to the fact that within the
dipole approximation, the bulk second-harmonic generation (SHG) in
centrosymmetric materials is identically zero. The SHG process can occur only at
the surface where the inversion symmetry is broken.

SSHG is particularly useful for studying the surfaces of centrosymmetric
materials. From the theoretical point of view, the calculation of the nonlinear
surface susceptibility tensor, $\boldsymbol{\chi}(-2\omega;\omega,\omega)$,
proceeds as follows. To mimic the semi-infinite system, one constructs a
supercell consisting of a finite slab of material plus a vacuum region. Both the
size of the slab and the vacuum region should be such that the value of
$\boldsymbol{\chi}(-2\omega;\omega,\omega)$ is well converged. One has to
include a cut function to decouple the two halves of the supercell in order to
obtain the value of $\boldsymbol{\chi}(-2\omega;\omega,\omega)$ for either half.
If the supercell itself is centrosymmetric, the value
$\boldsymbol{\chi}(-2\omega;\omega,\omega)$ is identically zero, thus the cut
function is of paramount importance.\cite{reiningPRB94, andersonPRB15} The cut
function can be generalized to one that is capable of obtaining the value of
$\boldsymbol{\chi}(-2\omega;\omega,\omega)$ for any part of the slab. The depth
within the slab for which $\boldsymbol{\chi}(-2\omega;\omega,\omega)$ is nonzero
can thus be obtained. One can also study how
$\boldsymbol{\chi}(-2\omega;\omega,\omega)$ goes to zero towards the middle of
the slab where the centrosymmetry of the material is restored.\cite{mejiaRMF04}
Therefore, for the surface of any centrosymmetric material we can find the
thickness of the layer where $\boldsymbol{\chi}(-2\omega;\omega,\omega)\ne 0$.

In this article, based on the above approach for the calculation of
$\boldsymbol{\chi}(-2\omega;\omega,\omega)$, we develop a model for the SH
radiation from the surface of a centrosymmetric material. We call this model the
three layer model, which considers that the SH conversion takes place in a thin
layer just below the surface of the material that lies under the vacuum region
and above the bulk of the material. Of course, one can replace the vacuum region
with any medium as long as it is not SH active. However, most of the
experimental setups for measuring the SH radiation take place in vacuum or air.
We develop the model and derive general expressions for the SH radiation for the
commonly used polarization combinations of the incoming and outgoing electric
fields. We particularize the results for the (111), (110), and (100) crystalline
surfaces of centrosymmetric materials.

This paper is organized as follows. In Sec. \ref{sec:threelayer}, we present the
relevant equations and theory that describe the SHG yield. In Sec.
\ref{sec:rcases}, we present the explicit expressions for each combination of
input and output polarizations for the (111), (110), and (100) surfaces.
Finally, we list our conclusions and final remarks in Sec.
\ref{sec:conclusions}.

%%%%%%%%%%%%%%%%%%%%%%%%%%%%%%%%%%%%%%%%%%%%%%%%%%%%%%%%%%%%%%%%%%%%%%%%%%%%%%%%
%%%%%%%%%%%%%%%%%%%%%%%%%%%%%%%%%%%%%%%%%%%%%%%%%%%%%%%%%%%%%%%%%%%%%%%%%%%%%%%%

\section{Three layer model for SSHG radiation}\label{sec:threelayer}

\begin{figure}[t]
\centering
\includegraphics[width=0.5\textwidth]{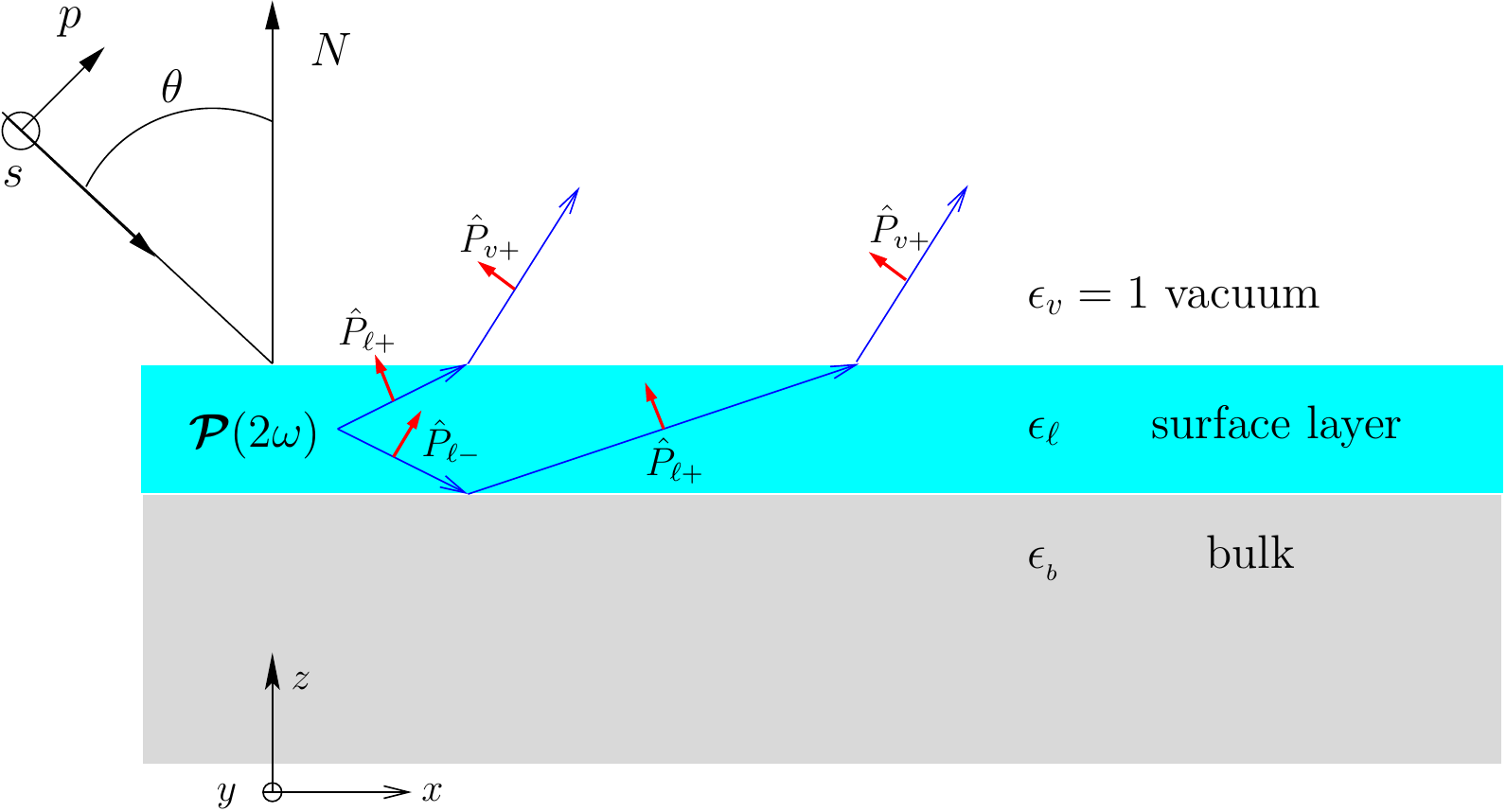}
\caption{Sketch of the three layer model for SHG. Vacuum is on top with
$\epsilon_v=1$, the layer with nonlinear polarization 
$\boldsymbol{\mathcal{P}}(2\omega)$
is characterized with $\epsilon_{\ell}(\omega)$ and the bulk with
$\epsilon_{b}(\omega)$. In the dipolar approximation the bulk does not radiate
SHG. The thin arrows are along the direction of propagation, and the unit
vectors for $p$-polarization are denoted with thick arrows (capital letters
denote SH components). The unit vector for $s$-polarization points along $-y$
(out of the page).\label{fig:3layer}}
\end{figure}

In this section we derive the formulas required for the calculation of the SSHG
yield, defined by
\begin{equation}\label{uno}
\mathcal{R}=\frac{I(2\omega)}{I^{2}(\omega)},
\end{equation}
with the intensity given by\cite{boyd}
\begin{equation}\label{dos}
I(\omega)=
\left\{
\begin{array}{cc}
\frac{c}{2\pi}n(\omega) |E(\omega)|^{2}
& \text{(cgs units)} \\\\
2\epsilon_{0}c\, n(\omega)|E(\omega)|^{2}
& \text{(MKS units)}
\end{array}
\right.,
\end{equation}
where $n(\omega)=\sqrt{\epsilon(\omega)}$ is the index of refraction with
$\epsilon(\omega)$ the dielectric function, $\epsilon_{0}$ is the vacuum
permittivity, and $c$ the speed of light in vacuum. We use Ref.
\onlinecite{mizrahiJOSA88} as a starting point for this work, as the derivation
of the three layer model is direct. In this scheme, we represent the surface by
three regions or layers. The first layer is the vacuum region (denoted by $v$)
with a dielectric function $\epsilon_{v}(\omega) = 1$ from where the fundamental
electric field $\mathbf{E}_{v}(\omega)$ impinges on the material. The second
layer is a thin layer (denoted by $\ell$) of thickness $d$ characterized by a
dielectric function $\epsilon_{\ell}(\omega)$. It is in this layer where the
second harmonic generation takes place. The third layer is the bulk region
denoted by $b$ and characterized by $\epsilon_{b}(\omega)$. Both the vacuum
layer and the bulk layer are semi-infinite (see Fig. \ref{fig:3layer}).

To model the electromagnetic response of the three-layer model we follow Ref.
\onlinecite{mizrahiJOSA88}, and assume a polarization sheet of the form
\begin{align}\label{m31}
\mathbf{P}(\mathbf{r}, t) = 
\boldsymbol{\mathcal{P}}e^{i\boldsymbol{\kappa}\cdot\mathbf{R}}
e^{-i\omega t}\delta(z - z_{\beta}) + \mathrm{c.c.},
\end{align}
where $\boldsymbol{\mathcal{P}}$ is the nonlinear polarization (given below), 
$\mathbf{R}=(x, y)$, $\boldsymbol{\kappa}$ is the component of the wave
vector $\boldsymbol{\nu}^{\strut}_\beta$ parallel to the surface, and
$z_{\beta}$ is the position of the sheet within medium $\beta$ (see Fig.
\ref{fig:3layer}). It is shown in Ref. \onlinecite{sipeJOSAB87} that the
solution of the Maxwell equations for the radiated fields $E_{\beta, p\pm}$ and
$E_{\beta, s}$, at points $z\neq 0$, with $\mathbf{P}(\mathbf{r}, t)$ acting as
a source can be written as
\begin{equation}\label{r2}
(E_{\beta, p\pm}, E_{\beta, s}) = 
 (\frac{\gamma i\tilde\omega^2}{\tilde w_{\beta}}
\,\hat{\mathbf{p}}_{\beta\pm}\cdot\boldsymbol{\mathcal{P}},
\frac{\gamma i\tilde\omega^2}{\tilde w_{\beta}}
\,\hat{\mathbf{s}}\cdot\boldsymbol{\mathcal{P}}),
\end{equation} 
where $\gamma=2\pi$ in cgs units and $\gamma=1/2\epsilon_0$ in MKS units.
$E_{\beta, p\pm}$ represents the electric field for $p$-polarization propagating
downward ($-$) or upward ($+$), and $E_{\beta, s}$ that for $s$-polarization,
both in medium $\beta$. Since for $s$-polarization the field is parallel to the
surface there is no need to distinguish the upward or downward direction of
propagation as it is needed for the $p$-polarized fields. Also,
$\tilde\omega=\omega/c$, and $\hat{\mathbf{s}}$ and
$\hat{\mathbf{p}}_{\beta\pm}$ are the unitary vectors for the $s$ and $p$
polarization of the radiated field, respectively. The $\pm$ notation refers to
upward ($+$) or downward ($-$) direction of propagation within medium $\beta$,
as shown in Fig. \ref{fig:3layer}. Thus,
\begin{equation}\label{r4}
\hat{\mathbf{p}}^{\strut}_{\beta\pm}(\omega) =
\frac{\kappa(\omega)\hat{\mathbf{z}}\mp
\tilde{w}_{\beta}(\omega)\hat{\boldsymbol{\kappa}}} 
{\tilde\omega n_{\beta}(\omega)}
=
\frac{\sin\theta_{0}\hat{\mathbf{z}}\mp 
w_{\beta}(\omega)\hat{\boldsymbol{\kappa}}} 
{n_{\beta}(\omega)}
,
\end{equation}
where $\kappa(\omega) =
|\boldsymbol{\kappa}(\omega)|=\tilde{\omega}\sin\theta_{0}$, 
$n_{\beta}(\omega) = \sqrt{\epsilon_{\beta}(\omega)}$ is the index of refraction
of medium $\beta$, and $z$ is the direction perpendicular to the surface that
points towards the vacuum. Lastly, $\tilde{w}_{\beta}(\omega)=\tilde{\omega}
w_{\beta}$, where
\begin{equation}\label{r3}
w^{\strut}_{\beta}(\omega) = 
\big(
\epsilon^{\strut}_{\beta}(\omega) - \sin^{2}\theta_{0}
\big)^{1/2},
\end{equation}
with $\theta_{0}$ the angle of incidence of $\mathbf{E}_{v}(\omega)$. We choose
the plane of incidence along the $\boldsymbol{\kappa}z$ plane, so
\begin{equation}\label{mc1}
\hat{\boldsymbol{\kappa}}
= \cos\phi\hat{\mathbf{x}} + \sin\phi\hat{\mathbf{y}},
\end{equation}
and
\begin{equation}\label{mmc2}
\hat{\mathbf{s}} = -\sin\phi\hat{\mathbf{x}} + \cos\phi\hat{\mathbf{y}},
\end{equation}
where $\phi$ is the azimuthal angle with respect to the $x$ axis.

In the three layer model, the nonlinear polarization responsible for the SSHG is
immersed in the thin $\beta=\ell$ layer, and is given by
\begin{equation}\label{tres}
\mathcal{P}_{\ell,i}(2\omega)=
\left\{
\begin{array}{cc}
  \chi_{ijk}(-2\omega;\omega,\omega)E_{\ell,j}(\omega)E_{\ell,k}(\omega)
& \text{(cgs units)} \\\\
  \epsilon_{0}\chi_{ijk}(-2\omega;\omega,\omega)E_{\ell,j}(\omega)E_{\ell,k}(\omega)
& \text{(MKS units)}
\end{array}
\right.,
\end{equation} 
where the tensor $\boldsymbol{\chi}(-2\omega;\omega,\omega)$ is the surface
nonlinear dipolar susceptibility  and the Cartesian indices $i,j,k$ are summed
over if repeated. 
We remark that the thickens of the layer $\ell$  is considered to be much
smaller than the wavelength of the fundamental field, thus
multiple reflections of both the fundamental and the SH can be neglected.
Also,
$\chi_{ijk}(-2\omega;\omega,\omega)=\chi_{ikj}(-2\omega;\omega,\omega)$ is the
intrinsic permutation symmetry due to the fact that SHG is degenerate in
$E_{\ell,j}(\omega)$ and $E_{\ell,k}(\omega)$. For ease of notation, we drop the
frequency argument from $\boldsymbol{\chi}(-2\omega;\omega,\omega)$ and we
simply write $\boldsymbol{\chi}$ from now on. As it was done in Ref.
\onlinecite{mizrahiJOSA88}, in presenting the results Eq.
\eqref{r2}-\eqref{mmc2} we have taken the polarization sheet (Eq. \eqref{m31})
to be oscillating at some frequency $\omega$. However, in the following we find
it convenient to use $\omega$ exclusively to denote the fundamental frequency
and $\boldsymbol{\kappa}$ to denote the component of the incident wave vector
parallel to the surface. Then the nonlinear generated polarization is
oscillating at $\Omega= 2\omega$ and will be characterized by a wave vector
parallel to the surface $\mathbf{K}=2\boldsymbol{\kappa}$. We can carry over
Eqs. \eqref{m31}-\eqref{mmc2} simply by replacing the lowercase symbols
($\omega, \tilde{\omega}, \boldsymbol{\kappa}, n_{\beta}, \tilde{w}_{\beta},
w_{\beta}, \hat{\mathbf{p}}_{\beta\pm}, \hat{\mathbf{s}}$) with uppercase
symbols ($\Omega, \tilde{\Omega}, \mathbf{K}, N_{\beta}, \tilde{W}_{\beta},
W_{\beta}, \hat{\mathbf{P}}_{\beta\pm}, \hat{\mathbf{S}}$), all evaluated at
$2\omega$. We always have that $\hat{\mathbf{S}}=\hat{\mathbf{s}}$.

To describe the propagation of the SH field, we see from Fig. \ref{fig:3layer},
that it is refracted at the layer-vacuum interface ($\ell v$), and  reflected
from the layer-bulk ($\ell b$) and layer-vacuum ($\ell v$) interfaces, thus we
define
\begin{equation}\label{r5}
\mathbf{T}^{\ell v}
= \hat{\mathbf{s}}T_{s}^{\ell v}\hat{\mathbf{s}} 
+ \hat{\mathbf{P}}_{v+}T_{p}^{\ell v}\hat{\mathbf{P}}_{\ell +},
\end{equation}
as the tensor for transmission from the $\ell v$ interface,
\begin{equation}\label{r6}
\mathbf{R}^{\ell b}
= \hat{\mathbf{s}}R_{s}^{\ell b}\hat{\mathbf{s}}
+ \hat{\mathbf{P}}_{\ell +}R_{p}^{\ell b}\hat{\mathbf{P}}_{\ell -},
\end{equation} 
as the tensor of reflection from the $\ell b$ interface, and
\begin{equation}\label{r6b}
\mathbf{R}^{\ell v}
= \hat{\mathbf{s}}R_{s}^{\ell v}\hat{\mathbf{s}}
+ \hat{\mathbf{P}}_{\ell -}R_{p}^{\ell v}\hat{\mathbf{P}}_{\ell +},
\end{equation} 
as that from the $\ell v$ interface. The Fresnel factors in uppercase letters,
$T^{ij}_{s,p}$ and $R^{ij}_{s,p}$, are evaluated at $2\omega$ from the following
well known formulas,\cite{ mizrahiJOSA88}
\begin{align}
t_s^{ij}(\omega) &=
\frac{2w_{i}(\omega)}{w_{i}(\omega)+w_{j}(\omega)},\\
t_{p}^{ij}(\omega) &=
\frac{2w_{i}(\omega)\sqrt{\epsilon_{i}(\omega)\epsilon_j(\omega)}}
     {w_{i}(\omega)\epsilon_{j}(\omega)+w_{j}(\omega)\epsilon_{i}(\omega)},\\
r_s^{ij}(\omega) &=
\frac{w_{i}(\omega) - w_{j}(\omega)}
     {w_{i}(\omega) + w_{j}(\omega)},\\
r_{p}^{ij}(\omega) &=
\frac{w_{i}(\omega)\epsilon_{j}(\omega) - w_{j}\epsilon_{i}(\omega)}
     {w_{i}(\omega)\epsilon_{j}(\omega) + w_{j}(\omega)\epsilon_{i}(\omega)}. 
\end{align}
From these expressions one can show that,
\begin{align}\label{mf}
1 + r^{\ell b}_{s} &= t^{\ell b}_{s}\nonumber\\
1 + r^{\ell b}_{p}
&= \frac{n_b}{n_\ell} 
t^{\ell b}_{p} 
\nonumber\\ 
1 - r^{\ell b}_{p}
&= \frac{n_\ell}{n_b}
   \frac{w_{b}}{w_{\ell}}t^{\ell b}_{p}\\ 
t^{\ell v}_{s,p} &= \frac{w_{\ell}}{w_{v}}t^{v\ell}_{s,p}\nonumber
.
\end{align}

%%%%%%%%%%%%%%%%%%%%%%%%%%%%%%%%%%%%%%%%%%%%%%%%%%%%%%%%%%%%%%%%%%%%%%%%%%%%%%%%
%%%%%%%%%%%%%%%%%%%%%%%%%%%%%%%%%%%%%%%%%%%%%%%%%%%%%%%%%%%%%%%%%%%%%%%%%%%%%%%%

\subsection{SSHG Yield}\label{sec:yield}

As explained above, we neglect multiple reflections, and then
we obtain the total $2\omega$ radiated field by using Eqs. \eqref{r5},
\eqref{r6}, and \eqref{r6b},
\begin{equation*}\label{r7}
\mathbf{E}(2\omega)
= E_s(2\omega)
\left(
\mathbf{T}^{\ell v} + \mathbf{T}^{\ell v}\cdot\mathbf{R}^{\ell b}
\right)
\cdot\hat{\mathbf{s}}
+ E_{p+}(2\omega)\mathbf{T}^{\ell v}\cdot\hat{\mathbf{P}}_{\ell +}
 + E_{p-}(2\omega)\mathbf{T}^{\ell v}
\cdot\mathbf{R}^{\ell b}\cdot\hat{\mathbf{P}}_{\ell-}.
\end{equation*}
The first term is  the transmitted $s$-polarized field, the second one is the
reflected and then transmitted $s$-polarized field and the third and fourth
terms are the equivalent fields for $p$-polarization. The transmission is from
the layer into vacuum, and the reflection between the layer and the bulk. After
some simple algebra, we obtain
\begin{equation}\label{r8}
\mathbf{E}_{\ell}(2\omega) = \frac{\gamma i\tilde{\Omega}}{W_{\ell}}
\mathbf{H}_{\ell}\cdot\boldsymbol{\mathcal{P}}_\ell(2\omega),
\end{equation}
where,
\begin{equation}\label{r9}
\mathbf{H}_{\ell}
= \hat{\mathbf{s}}\,T_s^{\ell v}\left(1+R_s^{\ell b}\right)\hat{\mathbf{s}}
+ \hat{\mathbf{P}}_{v+}T_{p}^{\ell v}
\left(
\hat{\mathbf{P}}_{\ell +} +R_{p}^{\ell b}\hat{\mathbf{P}}_{\ell -}
\right). 
\end{equation}
The magnitude of the radiated SH field is given by
$E(2\omega)=\hat{\mathbf{e}}^{\mathrm{F}}\cdot\mathbf{E}_\ell(2\omega)$, where
$\hat{\mathbf{e}}^{\mathrm{F}}$ is the unit vector of the final polarization,
with $\mathrm{F}=S,P$, and then, $\hat{\mathbf{e}}^S=\hat{\mathbf{s}}$ and
$\hat{\mathbf{e}}^P=\hat{\mathbf{P}}_{v+}$. We expand the second term in
parenthesis of Eq. \eqref{r9} as
\begin{equation*}\label{m1}
\begin{split}
\hat{\mathbf{P}}_{\ell +} + R_{p}^{\ell b}\hat{\mathbf{P}}_{\ell -}
&= \frac{\sin\theta_{0}\hat{\mathbf{z}} - W_{\ell}\hat{\boldsymbol{\kappa}}}
        {N_{\ell}}
 + R_{p}^{\ell b}
   \frac{\sin\theta_{0}\hat{\mathbf{z}} + W_{\ell}\hat{\boldsymbol{\kappa}}}
        {N_{\ell}}
\\\nonumber
&= \frac{1}{N_{\ell}}
\left(
\sin\theta_{0}(1+R^{\ell b}_{p})\hat{\mathbf{z}}
- W_{\ell}(1-R^{\ell b}_{p})\hat{\boldsymbol{\kappa}} 
\right)
\\\nonumber 
&= \frac{T^{\ell b}_{p}}{N^{2}_{\ell}N_{b}}
\left(
  N^{2}_{b}\sin\theta_{0}\hat{\mathbf{z}} 
- N^{2}_{\ell}W_{b}\hat{\boldsymbol{\kappa}}
\right)
,
\end{split}
\end{equation*}
and rewrite Eq. \eqref{r8} as
\begin{equation}\label{r10}
E(2\omega) =
\frac{2\gamma i \omega}{cW_{\ell}}
\hat{\mathbf{e}}^{\mathrm{F}}
\cdot
\mathbf{H}_{\ell}
\cdot
\boldsymbol{\mathcal{P}}_\ell(2\omega) 
= \frac{2\gamma i \omega}{cW_{v}}
\mathbf{e}^{\,2\omega,\mathrm{F}}_{\ell}
\cdot\boldsymbol{\mathcal{P}}_\ell(2\omega),
\end{equation}
where
\begin{equation}\label{r12mm}
\mathbf{e}^{2\omega,\mathrm{F}}_{\ell} =
\hat{\mathbf{e}}^{\mathrm{F}}\cdot 
\Bigg[
\hat{\mathbf{s}}T_{s}^{v\ell}T_{s}^{\ell b}\hat{\mathbf{s}} + 
\hat{\mathbf{P}}_{v+}
\frac{T^{v\ell}_{p}T^{\ell b}_{p}}
     {N^{2}_{\ell}N_{b}}
\left(
  N^{2}_{b}\sin\theta_{0}\hat{\mathbf{z}}
- N^{2}_{\ell}W_{b}\hat{\boldsymbol{\kappa}}
\right)
\Bigg].
\end{equation}

In the three layer model the nonlinear polarization is located in layer
$\ell$, thus, we evaluate the fundamental field required in Eq. \eqref{tres}
in this layer as well. We write
\begin{equation}\label{m2}
\begin{split}
\mathbf{E}_{\ell}(\omega)=E_0\left(
\hat{\mathbf{s}} t^{v\ell}_s(1+r^{\ell b}_s)\hat{\mathbf{s}}
+
\hat{\mathbf{p}}_{\ell-}
 t^{v\ell}_{p}
\hat{\mathbf{p}}_{v-}
+
\hat{\mathbf{p}}_{\ell+}
t^{v\ell}_{p}r^{\ell b}_{p}
\hat{\mathbf{p}}_{v-}
\right)\cdot\hat{\mathbf{e}}^{\mathrm{in}}=E_0\mathbf{e}^\omega_{\ell}
,
\end{split}
\end{equation} 
and following the steps that lead to Eq. \eqref{r12mm}, we find that
\begin{equation}\label{m12}
\mathbf{e}^{\omega,\mathrm{i}}_{\ell}
= \left[
\hat{\mathbf{s}}t_{s}^{v\ell}t_{s}^{\ell b}\hat{\mathbf{s}} 
+ \frac{t^{v\ell}_{p}t^{\ell b}_{p}}
       {n^{2}_{\ell}n_{b}}
\left(
  n^{2}_{b}\sin\theta_{0}\hat{\mathbf{z}}
+ n^{2}_{\ell}w_{b}\hat{\boldsymbol{\kappa}}
\right)
\hat{\mathbf{p}}_{v-}
\right]
\cdot\hat{\mathbf{e}}^{\mathrm{i}}.
\end{equation}  
Replacing $\mathbf{E}(\omega)\to E_0\mathbf{e}^{\omega,\mathrm{i}}_\ell$,  
in Eq. \eqref{tres}, we obtain that
\begin{equation}\label{m4}
\boldsymbol{\mathcal{P}}_\ell(2\omega) = 
\left\{
\begin{array}{cc}  
E^{2}_{0}\,\boldsymbol{\chi}:
\mathbf{e}^{\omega,\mathrm{i}}_{\ell}\mathbf{e}^{\omega,\mathrm{i}}_{\ell}
& \text{(cgs units)} \\\\
\epsilon_{0}E^{2}_{0}\,\boldsymbol{\chi}:
\mathbf{e}^{\omega,\mathrm{i}}_{\ell}\mathbf{e}^{\omega,\mathrm{i}}_{\ell}
& \text{(MKS units)} \\
\end{array}
\right.,
\end{equation}
where $\mathbf{e}^{\omega,\mathrm{i}}_{\ell}$ is given by Eq. \eqref{m12},
and thus
Eq. \eqref{r10} reduces to ($W_{v}=\cos\theta_{0}$)
\begin{equation}\label{mr10}
E(2\omega) =
\frac{2\eta i \omega}{c\cos\theta_{0}}
\mathbf{e}^{2\omega,\mathrm{F}}_{\ell}\cdot\boldsymbol{\chi}:
\mathbf{e}^{\omega,\mathrm{i}}_{\ell}\mathbf{e}^{\omega,\mathrm{i}}_{\ell}
,
\end{equation}
where $\eta=2\pi$ for cgs units and $\eta=1/2$ for MKS units. For ease of
notation, we define
\begin{equation}\label{mc0}
\Upsilon_{\mathrm{iF}}
\equiv 
\mathbf{e}^{2\omega,\mathrm{F}}_{\ell}\cdot\boldsymbol{\chi}:
\mathbf{e}^{\omega,\mathrm{i}}_{\ell}\mathbf{e}^{\omega,\mathrm{i}}_{\ell}
.
\end{equation}
From Eqs. \eqref{uno},
\eqref{dos}, and \eqref{mr10} we obtain that
\begin{equation}\label{mc6}
\mathcal{R}_{\mathrm{iF}}
=\frac{\eta\omega^{2}}{c^{3}\cos^{2}\theta_{0}}
\left\vert  
\frac{1}{n_{\ell}}
\Upsilon_{\mathrm{iF}}
\right\vert^{2} 
,
\end{equation}
as the SSHG yield, where $\eta =32\pi^3$ for cgs units and
$\eta=1/(2\epsilon_0)$ in MKS units. Since $\boldsymbol{\chi}$ is a surface
second order nonlinear susceptibility, in the MKS unit system is given in
$\mathrm{m}^{2}/\mathrm{V}$, and thus $\mathcal{R}_{\mathrm{iF}}$ is given in
$\mathrm{m}^{2}/\mathrm{W}$.

%%%%%%%%%%%%%%%%%%%%%%%%%%%%%%%%%%%%%%%%%%%%%%%%%%%%%%%%%%%%%%%%%%%%%%%%%%%%%%%%
%%%%%%%%%%%%%%%%%%%%%%%%%%%%%%%%%%%%%%%%%%%%%%%%%%%%%%%%%%%%%%%%%%%%%%%%%%%%%%%%

\section{\texorpdfstring{$\mathcal{R}_{\mathrm{iF}}$}{R} for different
polarization cases}\label{sec:rcases}

We obtain $\mathcal{R}_{\mathrm{iF}}$ from Eq. \eqref{mc6} for the most commonly
used polarizations of incoming and outgoing fields, i.e., iF=$pP$, $pS$, $sP$ or
$sS$. For this, we have to explicitly expand $\Upsilon_{\mathrm{iF}}$ (Eq.
\eqref{mc0}). First, by substituting Eqs. \eqref{mc1} and \eqref{mmc2} into Eq.
\eqref{r12mm}, we obtain
\begin{equation}\label{eq:e2wpmr}
\mathbf{e}^{2\omega,\mathrm{P}}_{\ell} =
\frac{T^{v\ell}_{p}T^{\ell b}_{p}}
     {N^{2}_{\ell}N_{b}}
\left(
  N^{2}_{b}\sin\theta_{0}\hat{\mathbf{z}}
- N^{2}_{\ell}W_{b}\cos\phi\hat{\mathbf{x}}
- N^{2}_{\ell}W_{b}\sin\phi\hat{\mathbf{y}}
\right),
\end{equation}
for $P$ $(\hat{\mathbf{e}}^{\mathrm{F}} = \hat{\mathbf{P}}_{v+})$ outgoing
polarization, and
\begin{equation}\label{eq:e2wsmr}
\mathbf{e}^{2\omega,\mathrm{S}}_{\ell} 
= T^{v\ell}_{s}T^{\ell b}_{s}
\left[-\sin\phi\hat{\mathbf{x}} + \cos\phi\hat{\mathbf{y}}\right].
\end{equation}
for $S$ $(\hat{\mathbf{e}}^{\mathrm{F}}=\hat{\mathbf{s}})$ outgoing
polarization. Secondly, using again Eqs. \eqref{mc1} and \eqref{mmc2}, but now
with Eq. \eqref{m12}, we obtain for $p$ incoming polarization
$(\hat{\mathbf{e}}^{\mathrm{i}} = \hat{\mathbf{p}}_{v-})$,
\begin{equation}\label{eq:ewewpmr}
\begin{split}
\mathbf{e}^{\omega,\mathrm{p}}_{\ell}\mathbf{e}^{\omega,\mathrm{p}}_{\ell}
= \left(\frac{t^{v\ell}_{p}t^{\ell b}_{p}}
{n^{2}_{\ell}n_{b}}\right)^{2}
\big(
  &n^{4}_{\ell}w^{2}_{b}\cos^{2}\phi
\hat{\mathbf{x}}\hat{\mathbf{x}}
+ 2n^{4}_{\ell}w^{2}_{b}\sin\phi\cos\phi
\hat{\mathbf{x}}\hat{\mathbf{y}}
+ 2n^{2}_{\ell}n^{2}_{b}w_{b}\sin\theta_{0}\cos\phi
\hat{\mathbf{x}}\hat{\mathbf{z}}\\
&+ n^{4}_{\ell}w^{2}_{b}\sin^{2}\phi
\hat{\mathbf{y}}\hat{\mathbf{y}}
+ 2n^{2}_{\ell}n^{2}_{b}w_{b}\sin\theta_{0}\sin\phi
\hat{\mathbf{y}}\hat{\mathbf{z}}
+ n^{4}_{b}\sin^{2}\theta_{0}
\hat{\mathbf{z}}\hat{\mathbf{z}}
\big),
\end{split}
\end{equation}
and for $s$ incoming polarization
$(\hat{\mathbf{e}}^{\mathrm{i}} = \hat{\mathbf{s}})$,
\begin{equation}\label{eq:ewewsmr}
\mathbf{e}^{\omega,\mathrm{s}}_{\ell}\mathbf{e}^{\omega,\mathrm{s}}_{\ell}
= \left(t^{v\ell}_{s}t^{\ell b}_{s}\right)^{2}
\left(
  \sin^{2}\phi\hat{\mathbf{x}}\hat{\mathbf{x}}
+ \cos^{2}\phi\hat{\mathbf{y}}\hat{\mathbf{y}} 
- 2\sin\phi\cos\phi\hat{\mathbf{x}}\hat{\mathbf{y}}
\right).
\end{equation}
So to calculate $\mathcal{R}_{\mathrm{iF}}$, we summarize in Table
\ref{tab:summary} the combination of the equations needed for all four
polarization cases. In the following subsections we write down the explicit
expressions for $\Upsilon_{\mathrm{iF}}$ for the most general case where the
surface has no symmetry other than that of noncentrosymmetry. We then develop
these expressions for particular cases of the most commonly investigated
surfaces, the (111), (100), and (110) crystallographic faces. For ease of
writing we split $\Upsilon_{\mathrm{iF}}$ as
\begin{equation}\label{mc25}
\Upsilon_{\mathrm{iF}} = \Gamma_{\mathrm{iF}}\,r_{\mathrm{iF}},
\end{equation} 
and in Table \ref{chis} we list,  for each surface, the components of
$\boldsymbol{\chi}$ different from 
zero.\cite{sipePRB87, popovbook}

\begin{table}[b]
\centering
\begin{tabular}{ | c | c | c | c | c | }
\hline
Case               & $\hat{\mathbf{e}}^{\mathrm{F}}$
                   & $\hat{\mathbf{e}}^{\mathrm{i}}$
                   & $\mathbf{e}^{2\omega,\mathrm{F}}_{\ell}$
                   & $\mathbf{e}^{\omega,\mathrm{i}}_{\ell}
                      \mathbf{e}^{\omega,\mathrm{i}}_{\ell}$ \\
\hline
$\mathcal{R}_{pP}$ & $\hat{\mathbf{P}}_{v+}$
                   & $\hat{\mathbf{p}}_{v-}$
                   &  Eq. \eqref{eq:e2wpmr} & Eq. \eqref{eq:ewewpmr} \\
$\mathcal{R}_{pS}$ & $\hat{\mathbf{S}}$
                   & $\hat{\mathbf{p}}_{v-}$
                   &  Eq. \eqref{eq:e2wsmr} & Eq. \eqref{eq:ewewpmr} \\
$\mathcal{R}_{sP}$ & $\hat{\mathbf{P}}_{v+}$
                   & $\hat{\mathbf{s}}$
                   &  Eq. \eqref{eq:e2wpmr} & Eq. \eqref{eq:ewewsmr} \\
$\mathcal{R}_{sS}$ & $\hat{\mathbf{S}}$
                   & $\hat{\mathbf{s}}$
                   &  Eq. \eqref{eq:e2wsmr} & Eq. \eqref{eq:ewewsmr} \\
\hline
\end{tabular}
\caption{Polarization unit vectors for $\hat{\mathbf{e}}^{\mathrm{F}}$ and
$\hat{\mathbf{e}}^{\mathrm{i}}$, and equations describing
$\mathbf{e}^{2\omega,\mathrm{F}}_{\ell}$ and
$\mathbf{e}^{\omega,\mathrm{i}}_{\ell}\mathbf{e}^{\omega,\mathrm{i}}_{\ell}$ for
each polarization case.\label{tab:summary}}
\end{table}

\begin{table}[t]
\begin{tabular}{|c|c|c|}
\hline 
(111)-$C_{3v}$     & (110)-$C_{2v}$  & (100)-$C_{4v}$ \\
\hline 
$\chi_{zzz}$ & $\chi_{zzz}$ & $\chi_{zzz}$\\
$\chi_{zxx}=\chi_{zyy}$ & $\chi_{zxx}\ne\chi_{zyy}$ & $\chi_{zxx}=\chi_{zyy}$\\
$\chi_{xxz}=\chi_{yyz}$ & $\chi_{xxz}\ne\chi_{yyz}$ & $\chi_{xxz}=\chi_{yyz}$\\
$\chi_{xxx}=-\chi_{xyy}=-\chi_{yyx}$ & &  \\
\hline 
\end{tabular}
\caption{Components of $\boldsymbol{\chi}$ for the (111), (110) and
  (100) crystallographic faces, belonging to the 
$C_{3v}$, 
$C_{2v}$, and
$C_{4v}$, symmetry groups, respectively. 
For the (111) surface we choose the $x$ and $y$ axes along 
the [$11\bar{2}$] and [$1\bar{1}0$] directions, respectively.
For the (110) and (100) we consider the $y$ axis perpendicular to the
plane of symmetry.\cite{sipePRB87}
We remark that in general
$\boldsymbol{\chi}^{(111)}\ne \boldsymbol{\chi}^{(110)}
\ne \boldsymbol{\chi}^{(100)}$.
}
\label{chis}
\end{table}

%%%%%%%%%%%%%%%%%%%%%%%%%%%%%%%%%%%%%%%%%%%%%%%%%%%%%%%%%%%%%%%%%%%%%%%%%%%%%%%%
%%%%%%%%%%%%%%%%%%%%%%%%%%%%%%%%%%%%%%%%%%%%%%%%%%%%%%%%%%%%%%%%%%%%%%%%%%%%%%%%

\subsection{\texorpdfstring{$\mathcal{R}_{pP}$}{RpP}}\label{sec:RpP} 

Per Table \ref{tab:summary}, $\mathcal{R}_{pP}$ requires Eqs. \eqref{eq:e2wpmr}
and \eqref{eq:ewewpmr}. After some algebra, we obtain that
\begin{equation}\label{mc78}
\Gamma_{pP} =
\frac{T^{v\ell}_{p}T^{\ell b}_{p}}
     {N^{2}_{\ell}N_{b}}
\left(\frac{t^{v\ell}_{p}t^{\ell b}_{p}}
{n^{2}_{\ell}n_{b}}\right)^{2}.
\end{equation}
and
\begin{equation}
\begin{split}
r_{pP}
=
- N^{2}_{\ell}W_{b}\big(
&+ n^{4}_{\ell}w^{2}_{b}\cos^{3}\phi\chi_{xxx}
 + 2n^{4}_{\ell}w^{2}_{b}\sin\phi\cos^{2}\phi\chi_{xxy}
 + 2n^{2}_{b}n^{2}_{\ell}w_{b}\sin\theta_{0}\cos^{2}\phi\chi_{xxz}\\
&+ n^{4}_{\ell}w^{2}_{b}\sin^{2}\phi\cos\phi\chi_{xyy}
 + 2n^{2}_{b}n^{2}_{\ell}w_{b}\sin\theta_{0}\sin\phi\cos\phi\chi_{xyz}
 + n^{4}_{b}\sin^{2}\theta_{0}\cos\phi\chi_{xzz}
  \big)\\
%%%%%%%%%%%%%%%%%%%%%%%%%%%%%
- N^{2}_{\ell}W_{b}\big(
&+ n^{4}_{\ell}w^{2}_{b}\sin\phi\cos^{2}\phi\chi_{yxx}
 + 2n^{4}_{\ell}w^{2}_{b}\sin^{2}\phi\cos\phi\chi_{yxy}
 + 2n^{2}_{b}n^{2}_{\ell}w_{b}\sin\theta_{0}\sin\phi\cos\phi\chi_{yxz}\\
&+ n^{4}_{\ell}w^{2}_{b}\sin^{3}\phi\chi_{yyy}
 + 2n^{2}_{b}n^{2}_{\ell}w_{b}\sin\theta_{0}\sin^{2}\phi\chi_{yyz}
 + n^{4}_{b}\sin^{2}\theta_{0}\sin\phi\chi_{yzz}
  \big)\\
%%%%%%%%%%%%%%%%%%%%%%%%%%%%%%
+ N^{2}_{b}\sin\theta_{0}\big(
&+ n^{4}_{\ell}w^{2}_{b}\cos^{2}\phi\chi_{zxx}
 + 2n^{4}_{\ell}w^{2}_{b}\sin\phi\cos\phi\chi_{zxy}
 + n^{4}_{\ell}w^{2}_{b}\sin^{2}\phi\chi_{zyy}\\
&+ 2n^{2}_{\ell}n^{2}_{b}w_{b}\sin\theta_{0}\cos\phi\chi_{zzx}
 + 2n^{2}_{\ell}n^{2}_{b}w_{b}\sin\theta_{0}\sin\phi\chi_{zzy}
 + n^{4}_{b}\sin^{2}\theta_{0}\chi_{zzz}
  \big),
\end{split}
\end{equation}
where all 18 independent components of $\boldsymbol{\chi}$ valid for a surface
with no symmetries contribute to $\mathcal{R}_{pP}$. Recall that
$\chi_{ijk}=\chi_{ikj}$. Using Table \ref{chis}, we present the expressions for
each of the three surfaces being considered here. For the (111) surface we
obtain
\begin{equation}\label{rpp111}
r^{(111)}_{pP} = 
N^{2}_{b}\sin\theta_{0}(n^{4}_{b}\sin^{2}\theta_{0}\chi_{zzz} 
+ n^{4}_{\ell}w^{2}_{b}\chi_{zxx}) - n^{2}_{\ell}N^{2}_{\ell}w_{b}W_{b}(2n^{2}_{b}\sin\theta_{0}\chi_{xxz}
 + n^{2}_{\ell}w_{b}\chi_{xxx}\cos3\phi).
\end{equation}
where the three-fold azimuthal symmetry of the SHG signal, typical of the
$C_{3v}$ symmetry group, is seen in the $3\phi$ argument of the cosine function.
For the (110) we have that
\begin{equation}\label{eq:final-rpp.mr.110}
\begin{split}
r^{(110)}_{pP} &= 
N^{2}_{b}\sin\theta_{0}
\bigg[
n^{4}_{b}\sin^{2}\theta_{0}\chi_{zzz}
+ n^{4}_{\ell}w^{2}_{b}
\bigg(
\frac{\chi_{zyy} + \chi_{zxx}}{2} + \frac{\chi_{zyy} - \chi_{zxx}}{2}\cos2\phi
\bigg)
\bigg]\\
&\qquad- 2n^{2}_{b}n^{2}_{\ell}N^{2}_{\ell}w_{b}W_{b}\sin\theta_{0}
\left(
\frac{\chi_{yyz} + \chi_{xxz}}{2} + \frac{\chi_{yyz} - \chi_{xxz}}{2}\cos2\phi
\right).
\end{split}
\end{equation}
The two-fold azimuthal symmetry of the SHG signal, typical of the $C_{2v}$
symmetry group, is seen in the $2\phi$ argument of the cosine function. For the
(100) surface we simply make $\chi_{zxx}=\chi_{zyy}$ and
$\chi_{xxz}=\chi_{yyz}$, as seen from Table \ref{chis}, and above expression
reduces to
\begin{equation}\label{rpp100}
r^{(100)}_{pP} = 
N^{2}_{b}\sin\theta_{0}\left(n^{4}_{b}\sin^{2}\theta_{0}\chi_{zzz}
+ n^{4}_{\ell}w^{2}_{b}\chi_{zxx}\right)
- 2n^{2}_{b}n^{2}_{\ell}N^{2}_{\ell}w_{b}W_{b}\sin\theta_{0}\chi_{xxz}.
\end{equation}
where we mention that the azimutal $4\phi$ symmetry for the $C_{4v}$ group of
the (100) surface is absent in above expresion since such contribution is only
related to the bulk nonlinear quadrupolar SH term,\cite{sipePRB87} that is
neglected in this work.

%%%%%%%%%%%%%%%%%%%%%%%%%%%%%%%%%%%%%%%%%%%%%%%%%%%%%%%%%%%%%%%%%%%%%%%%%%%%%%%%
%%%%%%%%%%%%%%%%%%%%%%%%%%%%%%%%%%%%%%%%%%%%%%%%%%%%%%%%%%%%%%%%%%%%%%%%%%%%%%%%

\subsection{\texorpdfstring{$\mathcal{R}_{pS}$}{RpS}}\label{sec:RpS}

Per Table \ref{tab:summary}, $\mathcal{R}_{pS}$ requires Eqs. \eqref{eq:e2wsmr}
and \eqref{eq:ewewpmr}. After some algebra, we obtain that
\begin{equation}\label{mcv}
\Gamma_{pS} =
T^{v\ell}_{s}T^{\ell b}_{s}\left(\frac{t^{v\ell}_{p}t^{\ell b}_{p}}
      {n^{2}_{\ell}n_{b}}\right)^{2}.
\end{equation}
and
\begin{equation}\label{eq:rpsfull}
\begin{split}
r_{pS}=
&- n^{4}_{\ell}w^{2}_{b}\sin\phi\cos^{2}\phi\chi_{xxx}
 - 2n^{4}_{\ell}w^{2}_{b}\sin^{2}\phi\cos\phi\chi_{xxy}
 - 2n^{2}_{\ell}n^{2}_{b}w_{b}\sin\theta_{0}\sin\phi\cos\phi\chi_{xxz}\\
&- n^{4}_{\ell}w^{2}_{b}\sin^{3}\phi\chi_{xyy}
 - 2n^{2}_{\ell}n^{2}_{b}w_{b}\sin\theta_{0}\sin^{2}\phi\chi_{xyz}
 - n^{4}_{b}\sin^{2}\theta_{0}\sin\phi\chi_{xzz}\\
%%%%%%%%%%%%%%%%%%%%%%%%%%%%%%%%%%%%%%%%%%%
&+ n^{4}_{\ell}w^{2}_{b}\cos^{3}\phi\chi_{yxx}
 + 2n^{4}_{\ell}w^{2}_{b}\sin\phi\cos^{2}\phi\chi_{yxy}
 + 2n^{2}_{\ell}n^{2}_{b}w_{b}\sin\theta_{0}\cos^{2}\phi\chi_{yxz}\\
&+ n^{4}_{\ell}w^{2}_{b}\sin^{2}\phi\cos\phi\chi_{yyy}
 + 2n^{2}_{\ell}n^{2}_{b}w_{b}\sin\theta_{0}\sin\phi\cos\phi\chi_{yyz}
 + n^{4}_{b}\sin^{2}\theta_{0}\cos\phi\chi_{yzz}
,
\end{split}
\end{equation}
In this case 12 out of the 18 components of $\boldsymbol{\chi}$ valid for a
surface with no symmetries, contribute to $\mathcal{R}_{pS}$. This is so,
because there is no $\mathcal{P}_{\ell,z}$ component, as the outgoing
polarization is $S$. From Table \ref{chis} we obtain,
\begin{equation}\label{eq:final-rps.111}
r^{(111)}_{pS} = - n^{4}_{\ell}w^{2}_{b}\chi_{xxx}\sin3\phi,
\end{equation}
for the (111) surface,
\begin{equation}\label{eq:final-rps.110}
r^{(110)}_{pS} = 
n^{2}_{\ell}n^{2}_{b}w_{b}\sin\theta_{0}(\chi_{yyz} - \chi_{xxz})\sin2\phi,
\end{equation}
for the (110) surface, 
finally,
\begin{equation}\label{r100ps}
r^{(100)}_{pS} = 0,
\end{equation}
for the (100) surface, where again, the zero value is only surface related as we
neglect  the bulk nonlinear quadrupolar contribution.

%%%%%%%%%%%%%%%%%%%%%%%%%%%%%%%%%%%%%%%%%%%%%%%%%%%%%%%%%%%%%%%%%%%%%%%%%%%%%%%%
%%%%%%%%%%%%%%%%%%%%%%%%%%%%%%%%%%%%%%%%%%%%%%%%%%%%%%%%%%%%%%%%%%%%%%%%%%%%%%%%

\subsection{\texorpdfstring{$\mathcal{R}_{sP}$}{RsP}}\label{sec:RsP}

Per Table \ref{tab:summary}, $\mathcal{R}_{sP}$ requires Eqs. \eqref{eq:e2wpmr}
and \eqref{eq:ewewsmr}. After some algebra, we obtain that
\begin{equation}\label{eq:gammasp}
\Gamma_{sP} =
\frac{T^{v\ell}_{p}T^{\ell b}_{p}}{N^{2}_{\ell}N_{b}}
\left(t^{v\ell}_{s}t^{\ell b}_{s}\right)^{2}.
\end{equation}
and
\begin{equation}\label{eq:rspfull}
\begin{split}
r_{sP} = 
&N^{2}_{\ell}W_{b}\big(
- \sin^{2}\phi\cos\phi\chi_{xxx}
+ 2\sin\phi\cos^{2}\phi\chi_{xxy}
- \cos^{3}\phi\chi_{xyy}
  \big)\\
%%%%%%%%%%%%%%%%%%%%%%%%%%%%%%%%%%%%%%%%%%%%%%%%%%%%%%
+&N^{2}_{\ell}W_{b}\big(
- \sin^{3}\phi\chi_{yxx}
+ 2\sin^{2}\phi\cos\phi\chi_{yxy}
- \sin\phi\cos^{2}\phi\chi_{yyy}
  \big)\\
%%%%%%%%%%%%%%%%%%%%%%%%%%%%%%%%%%%%%%%%%%%%%%%%%%%%%%
+&N^{2}_{b}\sin\theta_{0}\big(
+ \sin^{2}\phi\chi_{zxx}
- 2\sin\phi\cos\phi\chi_{zxy}
+ \cos^{2}\phi\chi_{zyy}
  \big),
\end{split}
\end{equation}
In this case 9 out of the 18 components of $\boldsymbol{\chi}(2\omega)$ valid
for a surface with no symmetries, contribute to $\mathcal{R}_{sP}$. This is so,
because there is no $E_z(\omega)$ component, as the incoming polarization is
$s$. From Table \ref{chis} we get,
\begin{equation}\label{eq:final-rsp.111}
r^{(111)}_{sP} = 
N^{2}_{b}\sin\theta_{0}\chi_{zxx} + N^{2}_{\ell}W_{b}\chi_{xxx}\cos3\phi.
\end{equation}
for the (111) surface,
\begin{equation}\label{eq:final-rsp.110}
r^{(110)}_{sP} = 
N^{2}_{b}\sin\theta_{0}
\left(
\frac{\chi_{zxx} + \chi_{zyy}}{2} + \frac{\chi_{zyy} - \chi_{zxx}}{2}\cos2\phi
\right).
\end{equation}
for the (110) surface, and
\begin{equation}\label{eq:final-rsp.100}
r^{(100)}_{sP} = 
N^{2}_{b}\sin\theta_{0}\chi_{zxx}.
\end{equation}
for the (100) surface.

%%%%%%%%%%%%%%%%%%%%%%%%%%%%%%%%%%%%%%%%%%%%%%%%%%%%%%%%%%%%%%%%%%%%%%%%%%%%%%%%
%%%%%%%%%%%%%%%%%%%%%%%%%%%%%%%%%%%%%%%%%%%%%%%%%%%%%%%%%%%%%%%%%%%%%%%%%%%%%%%%

\subsection{\texorpdfstring{$\mathcal{R}_{sS}$}{RsS}}\label{sec:RsS}

Per Table \ref{tab:summary}, $\mathcal{R}_{sS}$ requires Eqs. \eqref{eq:e2wsmr}
and \eqref{eq:ewewsmr}. After some algebra, we obtain that
\begin{equation}\label{eq:gammass}
\Gamma_{sS} =
T^{v\ell}_{s}T^{\ell b}_{s}\left(t^{v\ell}_{s}t^{\ell b}_{s}\right)^{2}.
\end{equation}
and
\begin{equation}
r_{sS} = 
 - \sin^{3}\phi\chi_{xxx}
 + 2\sin^{2}\phi\cos\phi\chi_{xxy}
 - \sin\phi\cos^{2}\phi\chi_{xyy}
 + \sin^{2}\phi\cos\phi\chi_{yxx}
 - 2\sin\phi\cos^{2}\phi\chi_{yxy}
 + \cos^{3}\phi\chi_{yyy}
.
\end{equation}
In this case 6 out of the 18 components of $\boldsymbol{\chi}(2\omega)$ valid
for a surface with no symmetries, contribute to $\mathcal{R}_{sS}$. This is so,
because there is neither an $E_{z}(\omega)$ component, as the incoming
polarization is $s$, nor a $\mathcal{P}_{\ell,z}(2\omega)$ component, as the
outgoing polarization is $S$. From Table \ref{chis}, we get
\begin{equation}
r^{(111)}_{sS} = \chi_{xxx}\sin3\phi,
\end{equation}
for the (111) surface,
\begin{equation}
r^{(110)}_{sS} = 0,
\end{equation}
and
\begin{equation}
r^{(100)}_{sS} = 0,
\end{equation}
for the (110) and (100) surfaces, respectively, both being zero as the bulk
nonlinear quadrupolar contribution is not considered here.

%%%%%%%%%%%%%%%%%%%%%%%%%%%%%%%%%%%%%%%%%%%%%%%%%%%%%%%%%%%%%%%%%%%%%%%%%%%%%%%%
%%%%%%%%%%%%%%%%%%%%%%%%%%%%%%%%%%%%%%%%%%%%%%%%%%%%%%%%%%%%%%%%%%%%%%%%%%%%%%%%

\section{Conclusions}\label{sec:conclusions}

We have derived the complete expressions for the SSHG radiation using the three
layer model to describe the radiating system. Our derivation yields the full
expressions for the radiation that include all required components of
$\chi_{ijk}$, regardless of symmetry considerations. Thus, these expressions can
be applied to any surface symmetry. We also reduce them according to the most
commonly used surface symmetries, the (111), (110), and (100) cases.

%%%%%%%%%%%%%%%%%%%%%%%%%%%%%%%%%%%%%%%%%%%%%%%%%%%%%%%%%%%%%%%%%%%%%%%%%%%%%%%%
%%%%%%%%%%%%%%%%%%%%%%%%%%%%%%%%%%%%%%%%%%%%%%%%%%%%%%%%%%%%%%%%%%%%%%%%%%%%%%%%

\bibliographystyle{unsrt}
%\bibliography{/Users/sma/Dropbox/docs/academics/master}
\bibliography{ref}

\begin{thebibliography}{10}

\bibitem{chenPRL81}
C.~K. Chen, A.~R.~B. de~Castro, and Y.~R. Shen.
\newblock Surface-enhanced second-harmonic generation.
\newblock {\em Phys. Rev. Lett.}, 46(2):145--148, January 1981.

\bibitem{shenNAT89}
Y.~R. Shen.
\newblock Surface properties probed by second-harmonic and sum-frequency
  generation.
\newblock {\em Nature}, 337(6207):519--525, February 1989.

\bibitem{mcgilpOE94}
J.~F. McGilp, M.~Cavanagh, J.~R. Power, and J.~D. O'Mahony.
\newblock Probing semiconductor interfaces using nonlinear optical
  spectroscopy.
\newblock {\em Opt. Eng.}, 33(12):3895--3900, 1994.

\bibitem{bloembergenAPB99}
N.~Bloembergen.
\newblock Surface nonlinear optics: a historical overview.
\newblock {\em Appl. Phys. B-Lasers O.}, 68(3):289--293, 1999.

\bibitem{mcgilpSRL99}
J.~F. McGilp.
\newblock Second-harmonic generation at semiconductor and metal surfaces.
\newblock {\em Surf. Rev. Lett.}, 6(03n04):529--558, 1999.

\bibitem{lupkeSSR99}
G.~L{\"u}pke.
\newblock Characterization of semiconductor interfaces by second-harmonic
  generation.
\newblock {\em Surf. Sci. Rep.}, 35(3):75--161, 1999.

\bibitem{downerPSSA01}
M.~C. Downer, Y.~Jiang, D.~Lim, L.~Mantese, P.~T. Wilson, B.~S. Mendoza, and
  V.I. Gavrilenko.
\newblock Optical second harmonic spectroscopy of silicon surfaces, interfaces
  and nanocrystals.
\newblock {\em Phys. Status Solidi A}, 188(4):1371--1381, 2001.

\bibitem{downerSIA01}
M.~C. Downer, B.~S. Mendoza, and V.~I. Gavrilenko.
\newblock Optical second harmonic spectroscopy of semiconductor surfaces:
  advances in microscopic understanding.
\newblock {\em Surf. Interface Anal.}, 31(10):966--986, 2001.

\bibitem{reiningPRB94}
L.~Reining, R.~Del~Sole, M.~Cini, and J.~G. Ping.
\newblock Microscopic calculation of second-harmonic generation at
  semiconductor surfaces: {As/Si(111)} as a test case.
\newblock {\em Phys. Rev. B}, 50(12):8411--8422, 1994.

\bibitem{andersonPRB15}
S.~M. Anderson, N.~Tancogne-Dejean, B.~S. Mendoza, and V.~V{\'e}niard.
\newblock Theory of surface second-harmonic generation for semiconductors
  including effects of nonlocal operators.
\newblock {\em Phys. Rev. B}, 91(7):075302, February 2015.

\bibitem{mejiaRMF04}
J.~E. Mej{\'i}a, B.~S. Mendoza, and C.~Salazar.
\newblock Layer-by-layer analysis of second harmonic generation at a simple
  surface.
\newblock {\em Revista Mexicana de F\'{\i}sica}, 50(2):134--139, 2004.

\bibitem{boyd}
Robert~W. Boyd.
\newblock {\em Nonlinear Optics}.
\newblock AP, New York, 2007.

\bibitem{mizrahiJOSA88}
V.~Mizrahi and J.~E. Sipe.
\newblock Phenomenological treatment of surface second-harmonic generation.
\newblock {\em J. Opt. Soc. Am. B}, 5(3):660--667, 1988.

\bibitem{sipeJOSAB87}
J.~E. Sipe.
\newblock New {Green}-function formalism for surface optics.
\newblock {\em Journal of the Optical Society of America B}, 4(4):481--489,
  1987.

\bibitem{sipePRB87}
J.~E. Sipe, D.~J. Moss, and H.~M. van Driel.
\newblock Phenomenological theory of optical second- and third-harmonic
  generation from cubic centrosymmetric crystals.
\newblock {\em Phys. Rev. B}, 35(3):1129--1141, January 1987.

\bibitem{popovbook}
S.~V. Popov, Y.~P. Svirko, and N.~I. Zheludev.
\newblock {\em Susceptibility tensors for nonlinear optics}.
\newblock CRC Press, 1995.

\end{thebibliography}

\end{document}